\documentclass[aps,prb,10pt,twocolumn, superscriptaddress,floatfix,reprint,longbibliography]{revtex4-2}
\usepackage{hyperref}
\usepackage{graphicx}
\usepackage[utf8]{inputenc}
\usepackage{amsmath}
\usepackage{mathtools}
\usepackage{algorithm}
\usepackage{algpseudocode}
\usepackage{verbatim}
\usepackage{amssymb}
\usepackage{color}
\usepackage{mathtools}
\usepackage{bm}
\usepackage{stmaryrd}
\usepackage{subfigure}
\usepackage{tikz}
\usepackage{ifthen}
\usepackage{physics}
\usepackage{soul}
\usepackage[normalem]{ulem}
\hypersetup{
    colorlinks=true,
    linkcolor=blue,
    filecolor=gray,
    urlcolor=blue,
    citecolor=blue,
    }
\usetikzlibrary{calc}
\usepackage{cancel}

\newcommand{\mat}[1]{\bm{#1}}

\definecolor{olivegreen}{HTML}{808000}

\begin{document}

\date{\today}\title{Minimal pole representation and analytic continuation of matrix-valued correlation functions}
\author{Lei Zhang}
\affiliation{Department of Physics, University of Michigan, Ann Arbor, Michigan 48109, USA}
\author{Yang Yu}
\affiliation{Department of Physics, University of Michigan, Ann Arbor, Michigan 48109, USA}
\author{Emanuel Gull}
\affiliation{Department of Physics, University of Michigan, Ann Arbor, Michigan 48109, USA}

\begin{abstract}
%Motivation
We present a minimal pole method for analytically continuing matrix-valued imaginary frequency correlation functions to the real axis, enabling precise access to off-diagonal elements and thus improving the interpretation of self-energies and susceptibilities in quantum simulations.
%Problem
Traditional methods for matrix-valued analytic continuation tend to be either noise-sensitive or make ad-hoc positivity assumptions.
Our approach avoids these issues via the construction of a compact pole representation with shared poles through exponential fits, expanding upon prior work focused on scalar functions.
%Results
We test our method across various scenarios, including fermionic and bosonic response functions, with and without noise, and for both continuous and discrete spectra of real materials and model systems. Our findings demonstrate that this technique addresses the shortcomings of existing methodologies, such as artificial broadening and positivity violations. The paper is supplemented with a sample implementation in Python.
\end{abstract}

\maketitle
\section{Introduction}
Numerical simulations of finite-temperature field theories typically obtain response functions, such as single-particle Green's functions and susceptibilities, on the Matsubara axis. Such methods include Monte Carlo simulations of lattice \cite{Blankenbecler81,Pollet12} and  quantum impurity \cite{gull2011continuous} problems in condensed matter, quantum chromodynamics simulations in nuclear and high-energy physics \cite{Asakawa01,Tripolt19,Rothkopf20}, and the self-consistent solution of low-order diagrammatic equations \cite{Hedin65,Dahlen05,Phillips14,Yeh22} and embedding problems \cite{Georges96,Kotliar06,Kananenka15,Zgid17} of solids and molecules. Analysis, interpretation, and comparison to experiment of these correlation functions require an additional step of analytic continuation to the real axis.

Analytic continuation of imaginary axis data is difficult due to the ill-posed nature of the direct inversion \cite{Jarrell96}. Numerous numerical methods for its regularization have been proposed, including interpolation with Pad\'{e} \cite{Baker96,Vidberg77} and Nevanlinna functions~\cite{Fei21,Fei21_Cara}, the maximum entropy method \cite{Jarrell96,Creffield95,Beach00,Gunnarsson10,Bergeron16,Levy17,Kraberger17}, approximation with Bayesian approaches \cite{Rothkopf13,Rumetshofer19}, stochastic fitting methods \cite{Sandvik98,Mishchenko00,Fuchs10,Goulko17,Krivenko19,Shao23}, and sparse modeling \cite{Otsuki17} of the inverse problem.

Recent work \cite{zhang2024minimal} showed that compact pole representations with complex poles are particularly powerful in this context. Pole expansions are in wide use in materials simulations \cite{Lu14} and have been used for analytic continuation in Refs.~\cite{ying2022analytic,ying2022pole,Huang23,HuangLi23,HuangShang24}. The methods in Ref.~\cite{zhang2024minimal} now make it possible to systematically converge analytic continuations for typical response functions as data quality is increased.

Moreover, compact pole representations also facilitate the interpretation of Green's functions since they provide a decomposition into damped or undamped `quasiparticles'. The representation is both efficient for storing Green's functions, requiring a small list of poles and weights rather than a discretization on a frequency grid, and for performing many-body calculations, enabling the use of residue calculus for diagram integration \cite{Gazizova24}.

Pole representations are closely related to the decomposition of real-time Green's functions into sums of decaying complex exponentials which, when Fourier transformed, correspond to sums of poles. In the case of signal-processing, the recovery of poles from data was already discussed in the eighteenth century by Gaspard de Prony \cite{Prony1795}. In the context of real-time propagation, extensions and approximations of this type have recently been analyzed with a wide variety of methods \cite{Barthel09,Tian21,Reeves23,Kemper24,Park24,Takahashi24}.

Most physical problems, especially in the context of simulations of real materials and ordered phases such as superconductivity, require the analysis of multi-orbital Green's functions with off-diagonal elements \cite{Gull14,Reymbaut15,Reymbaut17,Kraberger17,Yue23} that do not correspond to probability densities. While methods such as those presented in Ref.~\cite{zhang2024minimal} can in principle be applied to diagonal and off-diagonal elements of Green's function matrices independently, such a procedure may not respect 
the Carath\'{e}odory \cite{Fei21_Cara} structure of these objects, which (for fermion Green's functions) requires the matrix-valued response function to be a positive semidefinite function anywhere in the upper half of the complex plane. Knowledge of this matrix-valued structure is a prerequisite for performing certain post-processing calculations such as the extraction of two-particle correlation functions, including the optical and magnetic susceptibility, and the calculation of the Green's function on the real axis via the Dyson equation from an analytically continued self-energy.

A pole expansion in terms of shared real or complex poles with orbital-dependent matrix-valued weights, given by the form
\begin{align}\label{eq:pole_repr_complex}
    \mat{G}(z) = \sum_{l=1}^M \frac{\mat{A}_l}{z - \xi_l},\; \Im \xi_l \leq 0,
\end{align}
which is valid throughout the upper half of the complex plane, is therefore desirable.

This paper presents a minimal pole method (MPM) for generating such a representation in terms of a minimal number of poles $M$ on or below the real axis using two approximations with damped exponential functions and an intermediate holomorphic mapping. We show results for fermionic and bosonic systems with discrete and continuous spectra, with and without noise. We also show applications to the self-energy and Green's function of self-consistent real-materials calculations.

The remainder of this paper proceeds as follows. In Sec.~\ref{sec:theory} we give an overview of the analytic properties of Green's functions, introduce exponential approximation methods, and present the matrix-valued ESPRIT algorithm for extracting shared poles. In Sec.~\ref{sec:method} we discuss the three algorithmic steps of the pole approximation, the optional restriction to poles on the real axis, the optional combination with the discrete Lehmann representation, and applications beyond analytic continuation. Finally, Secs.~\ref{sec:results} and ~\ref{sec:conclusions} show numerical results and conclusions, respectively.

\section{Theory}\label{sec:theory}
\subsection{Green's functions} \label{sec:green_analytic_structure}
We consider matrix-valued response functions of interacting quantum systems, such as the matrix-valued single-particle Green's function $\mat{G} (z)$ for $z \in \mathbb{C}$ which, for a finite system at inverse temperature $\beta$, is defined as~\cite{negele2018quantum,Mahan13,Stefanucci13,Coleman15}
\begin{align}\label{eq:green_func}
  [\mat{G} (z)]_{ij} \!=\! \frac{1}{\mathcal{Z}}\!\sum_{m, n}\! \frac{e^{-\beta E_m} \!\pm\! e^{-\beta E_n}}{z \! - \! (E_n \! - \! E_m)}\!\bra{m} d_i \ket{n} \! \bra{n} d_j^\dag \ket{m} .
\end{align}
Here the indices $i$ and $j$ enumerate orbitals, the indices $m$ and $n$ label eigenstates; \(E_m\in\mathbb{R}\) are eigenvalues corresponding to eigenstates \(\ket{m}\), \(\mathcal{Z} = \sum_m e^{-\beta E_m}\) is the partition function, \(\beta\) is the inverse temperature, and \(d_i\) (\(d_i^\dag\)) are annihilation (creation) operators for orbitals \(1 \leq i \leq n_{\rm orb}\). The positive and negative signs correspond to fermionic and bosonic systems, respectively.

Eq.~(\ref{eq:green_func}) shows that the function $\mat{G}(z)$ is generated by poles on the real axis and is analytic in both the upper and lower half of the complex plane. $\mat{G}$  corresponds to the retarded Green's function just above the real axis and to the advanced one just below \cite{Mahan13,negele2018quantum}.

When \(z\) is evaluated at the discrete Matsubara frequencies $i\omega_n=i\frac{(2 n+1)\pi}{\beta}$ (for fermions) and $i\omega_n=i\frac{2 n \pi}{\beta}$ (for bosons), $n$ integer, $\mat{G}(z)$ corresponds to the Matsubara Green's function.  Numerical simulations of finite-temperature field theories, such as Monte Carlo simulations for lattices \cite{Blankenbecler81} and impurities \cite{gull2011continuous} or calculations of self-consistent diagrammatic perturbation theories \cite{Yeh22}, obtain response functions on the imaginary axis.

The corresponding single-particle spectral function \cite{Mahan13,negele2018quantum} is defined via the retarded Green's function as
\begin{equation}\label{eq:spec_func}
  \mat{A}(\omega) = -\frac{1}{\pi} {\rm Im}\mat{G}^{\rm Ret}(\omega) = -\frac{1}{\pi} {\rm Im}[\mat{G}(\omega + i0^+)] \; .
\end{equation}
The spectral function is related to the single-particle excitation spectrum of the system as measured, for example, in photoemission spectroscopy and is often the prime object of interest in many-body calculations. It is connected to the Matsubara Green's function as
\begin{equation}\label{eq:Mat_spec}
    \mat{G}^{\rm Mat}(i\omega_n) = \int_{-\infty}^{+\infty} d \omega \frac{\mat{A}(\omega)}{i\omega_n - \omega} \; .
\end{equation}

\subsection{Prony methods}\label{sec:Prony}
Prony's interpolation method~\cite{Prony1795} is a signal-processing technique originally proposed to decompose a uniformly sampled real-time signal $f(t)$ into a sum of decaying oscillatory signals. Its scalar version constructs an interpolation of the form 
\begin{equation}\label{eq:prony_orig1}
  f(t) = \sum_{i} A_i e^{-\alpha_i t} \cos(\omega_i t + \phi_i) \; ,
\end{equation}
estimating the amplitude $A_i$, damping factor $\alpha_i$, angular frequency $\omega_i$ and phase $\phi_i$ of a damped sinusoidal signal. After incorporating these real-valued variables into complex-valued variables $R_i$ and $s_i$, the equation can be rewritten as
\begin{equation}\label{eq:prony_orig2}
  f(t) = \sum_{i = 1}^{M} R_i e^{s_i t} \;,
\end{equation}
where $M$ is a positive integer. Thus Prony's method decomposes functions into a sum of complex exponentials. In practice, Prony’s interpolation method  is numerically unstable~\cite{moitra2015super}.

Rather than an interpolation problem, we consider here the related  approximation problem: given $N$ values of $f(t)$ sampled uniformly on a finite interval $[a, b]$ and given a target precision $\varepsilon > 0$, 
 find a set of complex weights $R_i$ and complex nodes $z_i$ such that 
\begin{equation}\label{eq:prony_scalar}
  \left| f_k - \sum_{i=1}^{M} R_i z_i^k \right| \leq \varepsilon \; \text{ for any } 0\leq k \leq N-1 \; ,
\end{equation}
where $f_k = f(a + k\Delta t)$ and $z_i = e^{s_i \Delta t}$, with $\Delta t = (b - a)/(N - 1)$. 

We further consider a generalization of Eq.~(\ref{eq:prony_scalar}) to a matrix-valued form:
\begin{equation}\label{eq:esprit_matrix}
  \left\Vert \mat{f}_k - \sum_{i=1}^{M} \mat{R}_i z_i^k \right\Vert \leq \varepsilon \; \text{ for any } 0\leq k \leq N-1 \; ,
\end{equation}
where the norm $\Vert \cdot \Vert$ of a matrix $\mat{f}$ is defined as  $\max_{ij}|f_{ij}|$, both $\mat{f}_k$ and $\mat{R}_i$ are $n_{\rm orb} \times n_{\rm orb}$ matrices, and  all elements of $\mat{R}_i$ share the same node $z_i$.

Several numerical methods that aim to find the best estimates of $M$, $\mat{R}_i$ and $z_i$ from possibly noise-contaminated data $\mat{f}_k$ have been developed. Among those are the Prony approximation method~\cite{beylkin2005approximation,beylkin2010approximation}, the Matrix Pencil Method~\cite{hua1990matrix,sarkar1995using,potts2013parameter}, and the Estimation of Signal Parameters via Rotational Invariance Techniques (ESPRIT)~\cite{roy1989esprit,potts2013parameter}. All of these methods can be understood as Prony-like methods~\cite{potts2013parameter}. Due to its robustness to noise and its computational efficiency, we will employ a variant of ESPRIT~\cite{roy1989esprit,potts2013parameter} in this paper, rather than the Prony approximation used in Ref.~\cite{zhang2024minimal}.

\subsection{Matrix ESPRIT algorithm}\label{sec:Esprit}
The generalization of ESPRIT to matrix-valued systems follows the work presented in Ref.~\cite{ying2022pole} as a matrix-valued generalization of the scalar-valued Prony's method of Ref.~\cite{ying2022analytic}.
We summarize here the main aspects.

Instead of directly operating on matrix-valued objects, we first flatten the  $n_{\rm orb} \times n_{\rm orb}$ matrices $\mat{f}_k$ and $\mat{R}_i$ into column vectors $\vec{f}_k$ and $\vec{R}_i$ of size $n_{\rm orb}^2$. The generalized ESPRIT method then introduces an auxiliary parameter $L$, which is typically chosen between $N / 3$ and $N / 2$ to minimize variance~\cite{sarkar1995using}, and utilizes a singular value decomposition (SVD) of the $n_{\rm orb}^2 (N - L)$ by $L+1$  matrix
\begin{equation}\label{eq:hankel_matrix}
    \mat{H} =
    \begin{pmatrix}
        \vec{f}_0       & \vec{f}_1     & \cdots & \vec{f}_L \\
        \vec{f}_1       & \vec{f}_2     & \cdots & \vec{f}_{L+1} \\
        \vdots    & \vdots  & \ddots & \vdots \\
        \vec{f}_{N-L-1} & \vec{f}_{N-L} & \cdots & \vec{f}_{N-1}
    \end{pmatrix}
\end{equation}
as
\begin{equation}\label{eq:Hankel}
    \hspace{-5pt}
    \mat{H} = \mat{U} \mat{\Sigma} \mat{W},
\end{equation}
where $\mat{U}$, $\mat{W}$ are $n_{\rm orb}^2(N - L)$ by $n_{\rm orb}^2(N - L)$ and $L+1$ by $L+1$ unitary matrices, respectively, $\mat{\Sigma}$ is a rectangular diagonal matrix of size $n_{\rm orb}^2(N - L)$ by $L + 1$ with diagonal elements $\sigma_0 \geq \sigma_1 \geq \cdots \geq \sigma_{L} \geq 0$.

The number of nodes $M$ is estimated as the smallest number so that $\sigma_{M}$ is under the predetermined precision $\varepsilon$, which is either limited by the noise level of $\vec{f}$ or by machine precision. As illustrated in the attached example code, where $\varepsilon$ is unknown one may also try to automatically determine $\varepsilon$ and $\sigma_M$ based on the distribution of singular values. The truncation to the minimum number of relevant singular values ensures that a minimal number of exponentials is obtained, thereby regularizing the continuation problem. The nodes $z_i$ are then obtained as the eigenvalues of the matrix $\mat{F}$:
\begin{equation}\label{eq:esprit_Fm}
    \mat{F} = (\mat{W}_0^T)^+ \mat{W}_1^T \;,
\end{equation}
where $+$ denotes the pseudo-inverse, $T$ denotes the transpose, and $\mat{W}_0$ and $\mat{W}_1$ are obtained from the first $M$ rows of $\mat{W}$ by deleting the last and the first column, respectively. Explicitly, they can be expressed as
\begin{equation}\label{eq:esprit_Wm}
    \mat{W}_s = \mat{W}(1:M, 1+s:L+s), \; s=0,1.
\end{equation} 

Finally, the vectorized weights $\vec{R}_i$ are recovered by solving the overdetermined least-squares Vandermonde system
\begin{equation}\label{eq:solve_weights_matrix}
\begin{pmatrix}
    \vec{f}_0^T \\
    \vec{f}_1^T \\
    \vdots \\
    \vec{f}_{N-1}^T
\end{pmatrix}
=  
\begin{pmatrix}
1 & 1 & \cdots & 1 \\
z_1 & z_2 & \cdots & z_M \\
\vdots    & \vdots    & \ddots & \vdots \\
z_1^{N-1} & z_2^{N-1} & \cdots & z_M^{N-1}
\end{pmatrix}
\begin{pmatrix}
    \vec{R}_1^T \\
    \vec{R}_2^T \\
    \vdots \\
    \vec{R}_{M}^T
\end{pmatrix}
\end{equation}
and are reshaped to $n_{\rm orb} \times n_{\rm orb}$ matrices $\mat{R}_i$. 
In this way, all sampling data $\mat{f}_k$ are approximated  within the error tolerance $\varepsilon$ by the smallest number of exponentials. Varying $k$ continuously over $[0, N-1]$, yields an approximation over the continuous interval of the form of Eq.~\eqref{eq:prony_orig2}, 
\begin{equation}\label{eq:esprit_continuous}
    \mat{f}(t) \approx \sum_{i = 1}^M \mat{R}_i \exp{{\frac{\ln z_i}{\Delta t}(t - a)}} ,\;\; t\in[a,b] \; .
\end{equation}

This supplemental materials of this paper contain an open-source implementation of the algorithm, which was used to generate the results shown below. In all simulations, \( L \) is fixed to \( 2N / 5 \).

\section{Minimal Pole Method}\label{sec:method}
We aim to construct a pole approximation of the form given in Eq.~\eqref{eq:pole_repr_complex} in the upper half of the complex plane, based on data provided along the Matsubara axis. The evaluation of this function just above the real axis will then yield the analytically continued spectral function, see Eq.~\eqref{eq:spec_func}.

In analogy to the scalar-valued version of the method~\cite{zhang2024minimal}, the construction of the matrix-valued approximation consists of three steps. First, we approximate the Matsubara data on a finite interval of the imaginary axis using the scalar-valued ESPRIT method for each matrix element. Second, we map this interval onto the unit circle using the holomorphic mapping presented in Ref.~\cite{zhang2024minimal} and evaluate the moments of the approximated function numerically. Finally, we apply matrix-valued ESPRIT to extract a minimal number of shared complex poles, map the poles back using the corresponding inverse holomorphic mapping, and evaluate the spectral function.

We discuss these three steps in the following.

\subsection{Approximation on the Matsubara axis}\label{sec:firstProny}
Given input data \(\mat{G}_{\rm input}(i\omega_n)\) on a uniform grid along the non-negative Matsubara axis, we aim to find an approximation to the Green's function that is valid both at the input Matsubara grid points and in the intervals between them.

While many interpolation or fitting methods could be used to approximate this function, including a least squares fit to compact basis functions \cite{Boehnke11,shinaoka2017compressing,Gull18,Kaye22} or an interpolation with cubic splines \cite{Kananenka16}, a fit with the minimum number of decaying exponential functions results in a representation that is accurate both at the interpolation nodes and in between, since it minimizes spurious oscillations but does not overfit the data. For further details, refer to Fig.~\ref{fig:least_square_vs_first_prony} as an example, discussed later.

Although the matrix-valued ESPRIT will become necessary in Sec.~\ref{sec:method3}, sharing the same set of poles in Eq.~(\ref{eq:pole_repr_complex}) for each matrix element does not necessarily imply that they can be well-approximated by shared exponentials through Eq.~(\ref{eq:esprit_matrix}). Therefore, we independently apply the scalar-valued exponential fit to each matrix element of the input data, yielding an approximation \(\mat{G}_{\rm approx}^{(L)}(iy)\) for \(\omega_0 \leq y \leq \omega_{n_{\omega} - 1}\) with a fixed parameter \(L\). This ensures that  
\begin{equation}
    \left\Vert\mat{G}_{\rm approx}^{(L)}(i \omega_n) - \mat{G}_{\rm input}(i \omega_n)\right\Vert \leq \varepsilon
\end{equation}
at the fitted Matsubara points.

Additionally varying \(L\) provides a convenient way to explore the sensitivity of the approximant between the interpolation points. If 
\begin{equation}\label{eq:first_esprit_L1_L2}
    \left\Vert\mat{G}_{\rm approx}^{(L')}(i y) - \mat{G}_{\rm approx}^{(L)}(i y)\right\Vert \leq \varepsilon
\end{equation}
holds for $\omega_n < y < \omega_{n+1}$ for different values of $L$ and $L'$, the points are likely oversampled and the approximation on $[\omega_n, \omega_{n+1}]$ is assumed to be accurate; otherwise the approximation on this interval may not be reliable. Typically only the low frequencies are not approximated reliably.

We may use this property to establish a lower threshold Matsubara frequency $\omega_{n_0}$. Above $\omega_{n_0}$ the approximation is assumed to be accurate in the continuous interval $y \in [\omega_{n_0}, \omega_{n_\omega - 1}]$. In most of the continuous cases we tested, this lower threshold coincides with the zero Matsubara frequency, i.e., $\omega_{n_0}=\omega_{0}$. In practice, we use \( L = 2N / 5 \) and \( L' = N / 2 \) in all our implementations.

\begin{figure}[tbh]
    \centering
\includegraphics[width=1.0 \columnwidth]{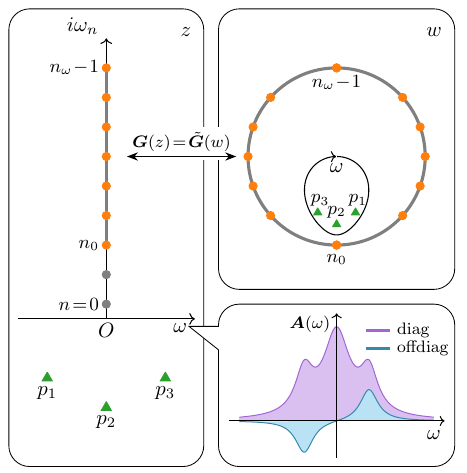}
\caption{Left: Complex plane with real and imaginary axis (black lines), Matsubara frequencies in the interval $[i\omega_{n_0}, i\omega_{n_\omega - 1}]$ (orange dots), remaining frequencies (grey dots), and pole locations of a spectral function (green triangles). Right top: Complex plane after holomorphic mapping (\ref{eq:con_map}) transforming the interval onto the unit circle, mapping the remainder of the complex plane to the interior of the unit disk, infinity to zero, and the poles to the interior of the egg-shaped image of the real axis. Bottom right: Example of diagonal and off-diagonal spectral functions belonging to shared complex poles (green triangles in other panels).}\label{fig:con_map}
\end{figure}

\subsection{Holomorphic mapping}\label{sec:method2}
The procedure of Sec.~\ref{sec:firstProny} results in an accurate description of the Green's function on an interval on the Matsubara axis, though not in the form of a pole representation. The problem now reduces to recovering the complex poles of such a function. As pointed out in Refs.~\cite{ying2022analytic,ying2022pole}, applying a holomorphic transform to map the interval to the unit circle reveals this pole information.

We choose the transform presented in Ref.~\cite{zhang2024minimal} to map the 
 entire complex plane into the closed unit disk \(\bar{D}\):
\begin{align}\label{eq:con_map}
  \left\{\begin{array}{lll}
    w\!\! &= g(z) &= z_{\rm s} - \sqrt{z_{\rm s}^2 + 1} \text{ with } z_{\rm s} = \frac{z-i\omega_{\rm m}}{\Delta \omega_{\rm h}}  \\
    z\!\! &= g^{-1}(w) \!\!&= \frac{\Delta \omega_{\rm h}}{2}(w - \frac{1}{w}) + i \omega_{\rm m}
  \end{array}
  \right. ,
\end{align}
where \(\omega_{\rm m} = (\omega_{n_0} + \omega_{n_\omega - 1}) / 2\), \(\Delta \omega_{\rm h} = (\omega_{n_\omega - 1} - \omega_{n_0}) / 2\), and the branch of the square root is chosen such that \(|w| \leq 1\) (cf. Fig.~\ref{fig:con_map}). This transformation offers several benefits: First, the correspondence is straightforward. Each point on the continuous interval \([i\omega_{n_0}, i\omega_{n_\omega - 1}]\) is mapped to two points with identical vertical coordinates, except for the endpoints \(i\omega_{n_0}\) and \(i\omega_{n_\omega - 1}\), which are mapped to the unique points \(-i\) and \(i\) on the unit circle, respectively. Second, it preserves the `left-right' symmetry $\Re g(x+iy) = -\Re g(-x + iy)$ of the plane, which simplifies the contour integrals that will be defined in Eq.~(\ref{eq:h_k_def}). Third,  every line segment \(dy\) on the continuous interval in the $z$-plane is mapped to the arc \((dx', dy')\) in the $w$-plane with \(dy' = \frac{1}{\Delta \omega_{\rm h}} dy\), where \(\Delta \omega_{\rm h}\), as defined above, is a constant. This avoids distortions of the interval along the vertical direction, which enhances robustness when some Matsubara points exhibit large fluctuations while others do not.

Eq.~(\ref{eq:con_map}) transforms the complex pole representation (\ref{eq:pole_repr_complex}) to:
\begin{equation}\label{eq:pole_repr_after_map}
    \tilde{\mat{G}}(w) = \sum_{l=1}^M \frac{\tilde{\mat{A}_l}}{w - \tilde{\xi}_l} + \text{analytic part} \;.
\end{equation}
With the matrix-valued integrals $\mat{h}_k$ over the unit circle defined as
\begin{equation}\label{eq:h_k_def}
    \mat{h}_k := \frac{1}{2\pi i} \int_{\partial \bar{D}}dw\tilde{\mat{G}}(w)w^k
\end{equation}
the residue theorem~\cite{ying2022analytic,ying2022pole} implies that:
\begin{equation}\label{eq:second_esprit}
    \mat{h}_k = \sum_{l=1}^M \tilde{\mat{A}}_l \tilde{\xi}_l^{k}, \; k \geq 0.
\end{equation}

The preservation of left-right symmetry in Eq.~(\ref{eq:con_map}) additionally 
simplifies the contour integral: 
\begin{align}\label{eq:hk_int}
  \mat{h}_k \!\!=\!\! 
  \left\{\begin{array}{lll}
    \!\!\!\!\!&\frac{i}{\pi} \int_{-\frac{\pi}{2}}^{\frac{\pi}{2}} \mat{G}(i (\omega_{\rm m} \! + \! \Delta \omega_{\rm h} \sin\theta)) \sin(k\!+\!1)\theta d \theta, \; k \text{ even} \\
    \!\!\!\!\!&\frac{1}{\pi} \int_{-\frac{\pi}{2}}^{\frac{\pi}{2}} \mat{G}(i (\omega_{\rm m} \! + \! \Delta \omega_{\rm h} \sin\theta)) \cos(k\!+\!1)\theta d \theta, \; k \text{ odd}
  \end{array}\right. \!\!\!.
\end{align}
Replacing \(\mat{G}(iy)\) with \(\mat{G}_{\rm approx}^{(L)}(iy)\) in Eq.~(\ref{eq:hk_int}) introduces uncertainty in \(\mat{h}_k\) on the order of the tolerance parameter \(\varepsilon\). In practice, each \(\mat{h}_k\) is calculated using an adaptive quadrature rule with error tolerance set much lower than \(\varepsilon\), and the number of integrals is truncated when \(\Vert \mat{h}_k \Vert < \varepsilon\). Since the holomorphic mapping maps poles into the interior of the unit disk, Eq.~(\ref{eq:second_esprit}) implies that the number of moments quickly decays as a function of $k$.

\subsection{Pole extraction}\label{sec:method3}
Eq.~(\ref{eq:second_esprit}) forms a second Prony problem (\ref{eq:esprit_matrix}). Applying the method of Sec.~\ref{sec:Esprit} to \(\mat{h}_k\) at precision \(\varepsilon\), the mapped pole weights \(\tilde{\mat{A}}_l\) and locations \(\tilde{\xi}_l\) are recovered. The exponentially decaying nature of the singular values of the $\mat{H}$~(\ref{eq:hankel_matrix}) ensures that a minimal number of complex poles, denoted by \(M\), are recovered in the current framework. 

The one-to-one correspondence between \(\tilde{\xi}_l\) and \(\xi_l\)~\cite{ying2022analytic,ying2022pole} via
\begin{align}\label{eq:location_trans}
\xi_l = g^{-1}(\tilde{\xi}_l) \; ,
\end{align}
is then used to recover the pole locations.

To recover the pole weights \(\mat{A}_l\), there are two main options, each with its own advantages. Given that the weights have a one-to-one correspondence via the residue theorem~\cite{zhang2024minimal}:
\begin{align}\label{eq:weight_trans}
    \mat{A}_l &= {\rm Res}[\mat{G}(z), \xi_l] \notag \\
    & = \left.\frac{d z}{d w}\right|_{\tilde{\xi}_l} \hspace{-6pt} \times {\rm Res}[\tilde{\mat{G}}(w), \tilde{\xi}_l] = \left.\frac{d z}{d w}\right|_{\tilde{\xi}_l} \hspace{-6pt} \times \tilde{\mat{A}}_l,
\end{align}
one approach is to recover \(\mat{A}_l\) directly from \(\tilde{\mat{A}}_l\). An advantage of this method is its effectiveness even when the input function contains a constant term, as seen in self-energy-like quantities, since the constant term can be automatically filtered out by Eq.~(\ref{eq:h_k_def}). Additionally, it is highly stable, as the noise has already been filtered in the first step. However, the transformation may introduce slight distortions, though these did not affect the overall shape of the spectrum in our tests (not shown). 

Alternatively, one may obtain the weights by solving the following system of equations:
\begin{equation}\label{eq:solve_pole_weights}
\!\!
\begin{pmatrix}
    \vec{G}_0^T \\
    \vec{G}_1^T \\
    \vdots \\
    \!\vec{G}_{n_\omega-1}^T\!
\end{pmatrix}
\!\!\!=\!\!\!
\begin{pmatrix}
\frac{1}{i\omega_0 - \xi_1} \!\!&\!\! \frac{1}{i\omega_0 - \xi_2} \!\!&\!\! \cdots \!\!&\!\! \frac{1}{i\omega_0 - \xi_M} \\
\frac{1}{i\omega_1 - \xi_1} \!\!&\!\! \frac{1}{i\omega_1 - \xi_2} \!\!&\!\! \cdots \!\!&\!\! \frac{1}{i\omega_1 - \xi_M} \\
\vdots    & \vdots    & \ddots & \vdots \\
\frac{1}{i\omega_{n_\omega - 1} - \xi_1} \!\!&\!\! \frac{1}{i\omega_{n_\omega - 1} - \xi_2} \!\!&\!\! \cdots \!\!&\!\!\! \frac{1}{i\omega_{n_\omega - 1} - \xi_M}
\end{pmatrix}\!\!\!
\begin{pmatrix}
    \vec{A}_1^T \\
    \vec{A}_2^T \\
    \vdots \\
    \vec{A}_{M}^T
\end{pmatrix},
\end{equation}
where \(\vec{G}_n\) and \(\vec{A}_l\) are the vectorized forms of \(\mat{G}_{\rm input}(i \omega_n)\) and \(\mat{A}_l\), as described in Sec.~\ref{sec:Esprit}. In this approach, no distortion occurs, and if some of the Matsubara points were ignored in the previous step ($n_0>0$) they can now be utilized. However, the method fails in the presence of a constant term. An optimization condition can also be incorporated into both methods to explicitly satisfy the analytic properties:
\begin{align}\label{eq:Al_real}
  \hspace{-5pt}
  \left.\begin{array}{ll}
    \mat{A}_l \text{ (for fermions) or } \text{sign}(\xi_l)\mat{A}_l \text{ (for bosons) is} \\
    \text{positive semidefinite with } (\sum_l \mat{A}_l)_{ij} = [d_i, d_j^\dag]_\pm
  \end{array}\right. ,
\end{align}
for discrete systems or 
\begin{align}\label{eq:Al_complex}
  \hspace{-5pt}
  \left.\begin{array}{ll}
    -\Im (\sum_l\frac{\mat{A}_l}{\omega - \xi_l}) \text{ (for fermions) or } -\text{sign}(\omega)\times \\
    \Im (\sum_l\frac{\mat{A}_l}{\omega - \xi_l}) \text{ (for bosons) is positive semide- }\\
    \text{finite for any } \omega \in \mathbb{R},
    \text{with } (\sum_l \mat{A}_l)_{ij} = [d_i, d_j^\dag]_\pm
  \end{array}\right.
\end{align}
for continuous systems, albeit at the cost of additional runtime due to the optimization process. As demonstrated in Ref.~\cite{zhang2024minimal}, even in the presence of noise, the non-causal parts of the unrestricted solutions are typically not observable. However, when the data is too noisy to reliably contain causal information, or when the spectrum exhibits singularities, imposing restrictions becomes crucial. 

In practice, we find that the second method yields better results for computing pole weights, so we always use Eq.~(\ref{eq:solve_pole_weights}). To address the issue of a non-zero constant term, we first calculate tentative weights \(\mat{A}_{l}^{\rm t}\) from the unrestricted solution of Eq.~(\ref{eq:weight_trans}). The constant term is then estimated as:
\begin{equation}
    {\rm const} = \frac{1}{n_{\omega} - n_0} \sum_{n=n_0}^{n_{\omega}-1}\left(\mat{G}_{\rm input}(i\omega_n) - \sum_{l=1}^M\frac{\mat{A}_l^{\rm t}}{i\omega_n - \xi_l}\right) \;.
\end{equation}
We then substitute \(\mat{G}_{\rm input}(i\omega_n) - {\rm const}\) into the left-hand side of Eq.~(\ref{eq:solve_pole_weights}) with optional restrictions (\ref{eq:Al_real}) or (\ref{eq:Al_complex}) to obtain the final weights \(\mat{A}_l\). After obtaining \(\mat{A}_l\) and \(\xi_l\), the spectral function is finally recovered using:
\begin{equation}\label{eq:spec_rec}
    \mat{A}_{\rm rec}(\omega) = -\frac{1}{\pi} \Im\left(\sum_{l=1}^M \frac{\mat{A}_l}{\omega + i 0 ^+ - \xi_l}\right) \; .
\end{equation}

\subsection{Restriction to real poles}
It is sometimes useful to restrict the locations of poles to the real axis, e.g., when it is known that the system consists of discrete excitations, or when a representation of the Green's function is desired that is valid both in the upper and in the lower half of the complex plane \cite{Gazizova24}. This restriction typically comes at the cost of a larger set of poles.

\begin{figure}[tbh]
    \centering
\includegraphics[width=1.0 \columnwidth]{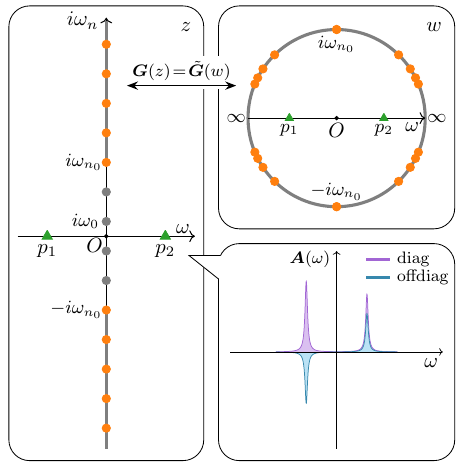}
\caption{Illustration of the holomorphic mapping (\ref{eq:con_map2}). Left: The original plane, showing Matsubara frequencies in the interval \((- \infty, -i\omega_{n_0}]\cup [i\omega_{n_0}, +\infty)\) (orange dots), remaining frequencies (grey dots), and pole locations of a spectral function (green triangles). Top right: The complex plane after the holomorphic mapping (\ref{eq:con_map2}), which transforms the interval onto the unit circle, maps the remainder of the complex plane inside the unit disk, and maps the real axis onto the line segment \((-1, 0)\) to \((1, 0)\), with the origin mapped to the origin and infinity mapped to \((\pm 1, 0)\). Bottom right: Examples of diagonal and off-diagonal spectral functions corresponding to the shared real poles. Delta peaks are broadened for better visualization.}\label{fig:con_map2}
\end{figure}

Instead of using Eq.~\eqref{eq:con_map}, now we consider the holomorphic mapping
\begin{align}\label{eq:con_map2}
  \left\{\begin{array}{lll}
    w\!\! &= g(z) &= \omega_{n_0}\left(\sqrt{\frac{1}{z^2} + \frac{1}{\omega_{n_0}^2}} - \frac{1}{z}\right) \\
    z\!\! &= g^{-1}(w) \!\!&= 2\omega_{n_0}\frac{w}{1-w^2}
  \end{array}
  \right. \; ,
\end{align}
which reduces to the holomorphic mapping presented in Ref.~\cite{ying2022analytic} for fermionic systems when \(n_0 = 0\).  This maps the continuous imaginary interval \((- \infty, -i\omega_{n_0}]\cup [i\omega_{n_0}, +\infty)\) to the unit circle, see Fig.~\ref{fig:con_map2} for more details. The  method of Sec.~\ref{sec:firstProny} provides a controlled approximation over the interval \([i\omega_{n_0}, i\omega_{n_\omega - 1}]\), while the tail behavior in \([i\omega_{n_\omega - 1}, +\infty)\) can be well approximated using other techniques, such as the complex pole representation introduced above. As for the approximation on \((- \infty, -i\omega_{n_0}]\), it can be directly derived from the symmetry $\mat{G}(z^*) = (\mat{G}(z))^\dag$.

The contour integral in Eq.~(\ref{eq:h_k_def}) can be simplified in this case to:
\begin{align}\label{eq:hk_int2}
  \!\!\mat{h}_k \!=\! 
  \left\{\begin{array}{lll}
    \!\!\!\!\!&\frac{i}{\pi} \int_{0}^{\frac{\pi}{2}} [\mat{G}(\frac{i \omega_{n_0}}{\sin \theta}) \!-\! (\mat{G}(\frac{i \omega_{n_0}}{\sin \theta}))^\dag] \sin(k\!+\!1)\theta d \theta, k \text{ even} \\
    \!\!\!\!\!&\frac{1}{\pi} \int_{0}^{\frac{\pi}{2}} [\mat{G}(\frac{i \omega_{n_0}}{\sin \theta}) \!+\! (\mat{G}(\frac{i \omega_{n_0}}{\sin \theta}))^\dag] \cos(k\!+\!1)\theta d \theta, k \text{ odd}
  \end{array}\right. \!\!.
\end{align}
In many cases, the Hamiltonian is real-valued, so that the Green's function exhibits the symmetry $[\mat{G}(z^*)]_{ij} = [\mat{G}^*(z)]_{ij}$. 
This symmetry further simplifies Eq.~(\ref{eq:hk_int2}) to:
\begin{align}\label{eq:hk_int3}
  \!\!\mat{h}_k \!=\! 
  \left\{\begin{array}{lll}
    \!\!\!\!\!&-\frac{2}{\pi} \int_{0}^{\frac{\pi}{2}} \Im[\mat{G}(\frac{i \omega_{n_0}}{\sin \theta})] \sin(k+1)\theta d \theta, \; k \text{ even} \\
    \!\!\!\!\!&+\frac{2}{\pi} \int_{0}^{\frac{\pi}{2}} \Re[\mat{G}(\frac{i \omega_{n_0}}{\sin \theta})] \cos(k+1)\theta d \theta, \; k \text{ odd}
  \end{array}\right. .
\end{align}

Since Eqs.~(\ref{eq:pole_repr_after_map}) to (\ref{eq:second_esprit}) still hold for the new mapping, all other steps presented in Sec.~\ref{sec:method} remain applicable, resulting in poles that lie on the real axis. In practice, unlike the generic method presented in the previous section, we found that obtaining pole weights from the transformed plane is more stable than from the original plane. Therefore, we  use Eq.~(\ref{eq:weight_trans}) rather than Eq.~(\ref{eq:solve_pole_weights}) in this case.

\subsection{Combination with the Discrete Lehmann Representation}\label{sec:dlr}
In situations where computational efficiency, rather than accuracy, is important, it is possible to utilize the discrete Lehmann representation (DLR)~\cite{Kaye22,kaye2022libdlr} directly for approximating Matsubara data. By specifying an error tolerance \(\varepsilon\) and a sufficient cutoff \(\Lambda = \beta \omega_{\rm max}\), where \(\omega_{\rm max}\) represents the energy support of the system and can be chosen to be sufficiently large, DLR  generates \(r\) fixed real poles \(\omega_i^{(\rm dlr)}\), where \(r = \mathcal{O}(\Lambda \log \frac{1}{\varepsilon})\). The pole weights \(\mat{g}_i^{(\rm dlr)}\) are then determined based on the input data \(\mat{G}_{\rm input}(i\omega_n)\). The resulting function, \(\mat{G}_{\rm approx}(i\omega_n) = \sum_{i=1}^r \frac{\mat{g}_i^{(\rm dlr)}}{i\omega_n - \omega_i^{(\rm dlr)}}\), provides an approximation of the Matsubara data with an accuracy within \(\varepsilon\).

Although the DLR method does not provide a precise approximation between Matsubara points (see Sec.~\ref{sec:approx_interval}), there typically is a threshold \(n_0\) such that it offers a precise approximation in the intervals \((-\infty, -i\omega_{n_0}] \cup [i\omega_{n_0}, +\infty)\). In practice, and especially at low temperatures, \(\omega_{n_0}\) is very close to zero, making the impact of disregarding the first \(n_0\) points negligible. In that case, the calculation of the contour integral in Eq.~(\ref{eq:h_k_def}), directly using the DLR expansion, simplifies to:
\begin{equation}\label{eq:h_k_dlr}
    \mat{h}_k = \sum_{i=1}^r \tilde{\mat{g}}_i^{({\rm dlr})} \tilde{\omega}_i^{({\rm dlr})k}, \; k \geq 0,
\end{equation}
where \(\tilde{\mat{g}}_i^{({\rm dlr})}\) and \(\tilde{\omega}_i^{({\rm dlr})}\) can be derived from \({\mat{g}}_i^{({\rm dlr})}\) and \({\omega}_i^{({\rm dlr})}\) using the transformations in Eqs.~(\ref{eq:weight_trans}) and (\ref{eq:location_trans}), respectively. This eliminates the need for time-consuming numerical quadrature. Additionally, there is no need to truncate \(\mat{h}_k\) at \(\Vert \mat{h}_k \Vert < \varepsilon\), since the number of recovered poles is strictly bounded by \(r\). The method described in Sec.~\ref{sec:method3} may then be used to recover the pole information.

The use of DLR rather than the first exponential approximation of Sec.~\ref{sec:firstProny} results in a loss of precision, especially at high temperature, and requires at least \(r\) input points. Results for this variant and an estimate of the loss of precision are shown in Sec.~\ref{sec:realistic}.

\subsection{Application to heating and cooling of self-consistent many-body simulations}
Apart from obtaining the spectral function, an additional application of this methodology consists in the temperature extrapolation of self-consistent many-body simulations \cite{Yu23}. In these methods, results that are obtained for one value of temperature are then used to initialize the convergence of self-consistent iterations at a different temperature. In the current context, this implies evaluating Eq.~\eqref{eq:pole_repr_complex} on the Matsubara grid for a different temperature and using those results to initialize self-consistent iterations. Such a procedure may be needed if multiple stable fixed-points exist, such as in the coexistence regime of a first-order transition. It may also be needed in situations where self-consistent iterations (in a damped, undamped, or accelerated \cite{Pokhilko22} convergence scheme) do not converge to a physical solution.

In these situations, the temperature extrapolation of the matrix-valued propagator results in a `physical' starting solution which inherits the spectra from a nearby temperature.  This procedure may substantially reduce the number of iterations needed for convergence, similar to the interpolation with Carath\'{e}odory functions used in Ref.~\cite{Yu23}.

\section{Results}\label{sec:results}
\subsection{Approximation on the imaginary frequency interval}\label{sec:approx_interval}
We begin the discussion of our results with an analysis of the first interpolation step described in Sec.~\ref{sec:firstProny}. A crucial insight is that the response function should be approximated accurately on the entire interpolation interval on the imaginary axis, not just on the Matsubara points, because of the subsequent integration steps [Eq.~\eqref{eq:hk_int} or Eq.\eqref{eq:hk_int2}]. In our method, this first approximation is obtained from ESPRIT. While the resulting approximant does not have the form of a pole representation, it forms a smooth function that can be evaluated anywhere on the Matsubara axis.

\begin{figure}[tbh]
    \centering
\includegraphics[width=1.0 \columnwidth]{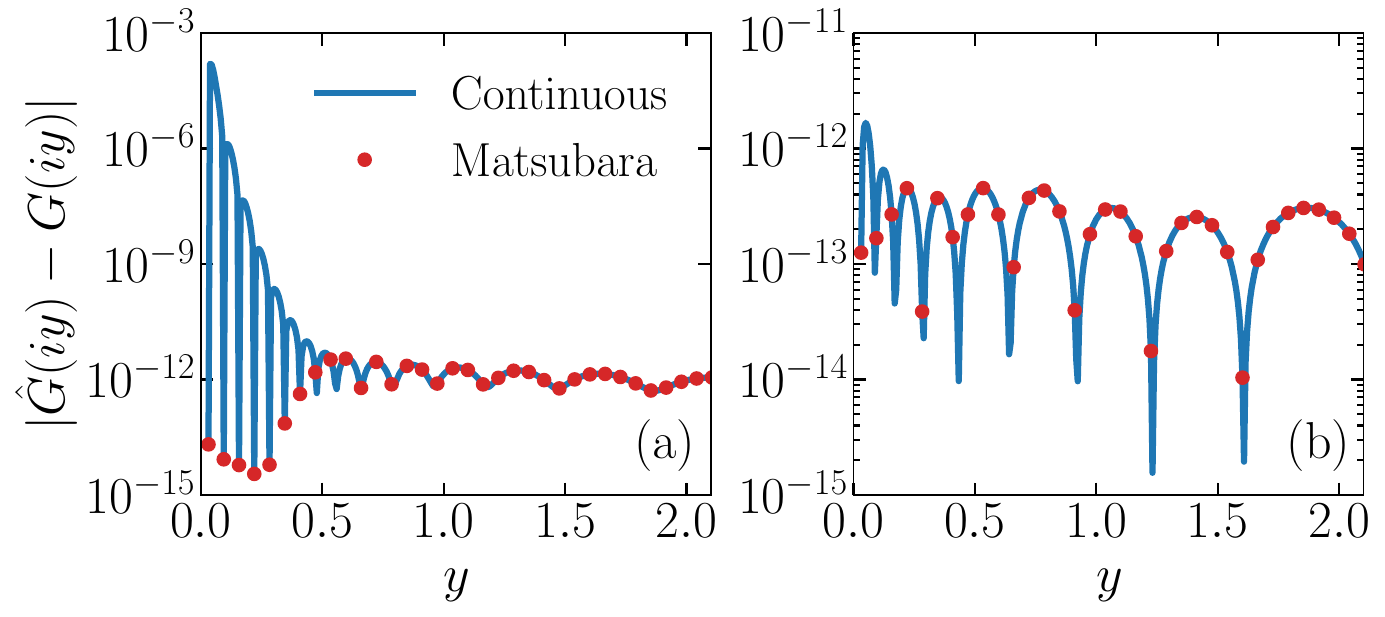}
\caption{Left panel: interpolation of an example Green's function at interpolation points and in between as obtained in a least squares fit to DLR coefficients. Right panel: same, as obtained with method of Sec.~\ref{sec:firstProny}.
} 
\label{fig:least_square_vs_first_prony}
\end{figure}

Figure~\ref{fig:least_square_vs_first_prony} compares different methods of fitting the Green’s function. The left panel shows a standard least squares fit to DLR coefficients, while the right panel uses the method described in Sec.~\ref{sec:Esprit}, both applied to the same Matsubara data. The model under consideration is a tight-binding model on the Bethe lattice with hopping $t$, for which the spectral function is a semicircle:
\begin{equation}
    A(\omega) = \frac{1}{2 \pi t^2} \sqrt{4t^2 - \omega^2} \; ,
\end{equation}
and the Green's function has the analytical form:
\begin{equation}
    G(iy) = \frac{i}{2t^2}(y - \sqrt{y^2 + 4 t^2}) \quad \text{for } y > 0 \; .
\end{equation}
The comparison is carried out with $t=1$ at $\beta = 100$. It is evident that ESPRIT (right panel) produces an accurate approximant both at the interpolation points and in between, unlike a standard least squares fit to DLR coefficients which results in large errors at small $y$ (left panel).

\begin{figure}[tbh]
    \centering
\includegraphics[width=1.0 \columnwidth]{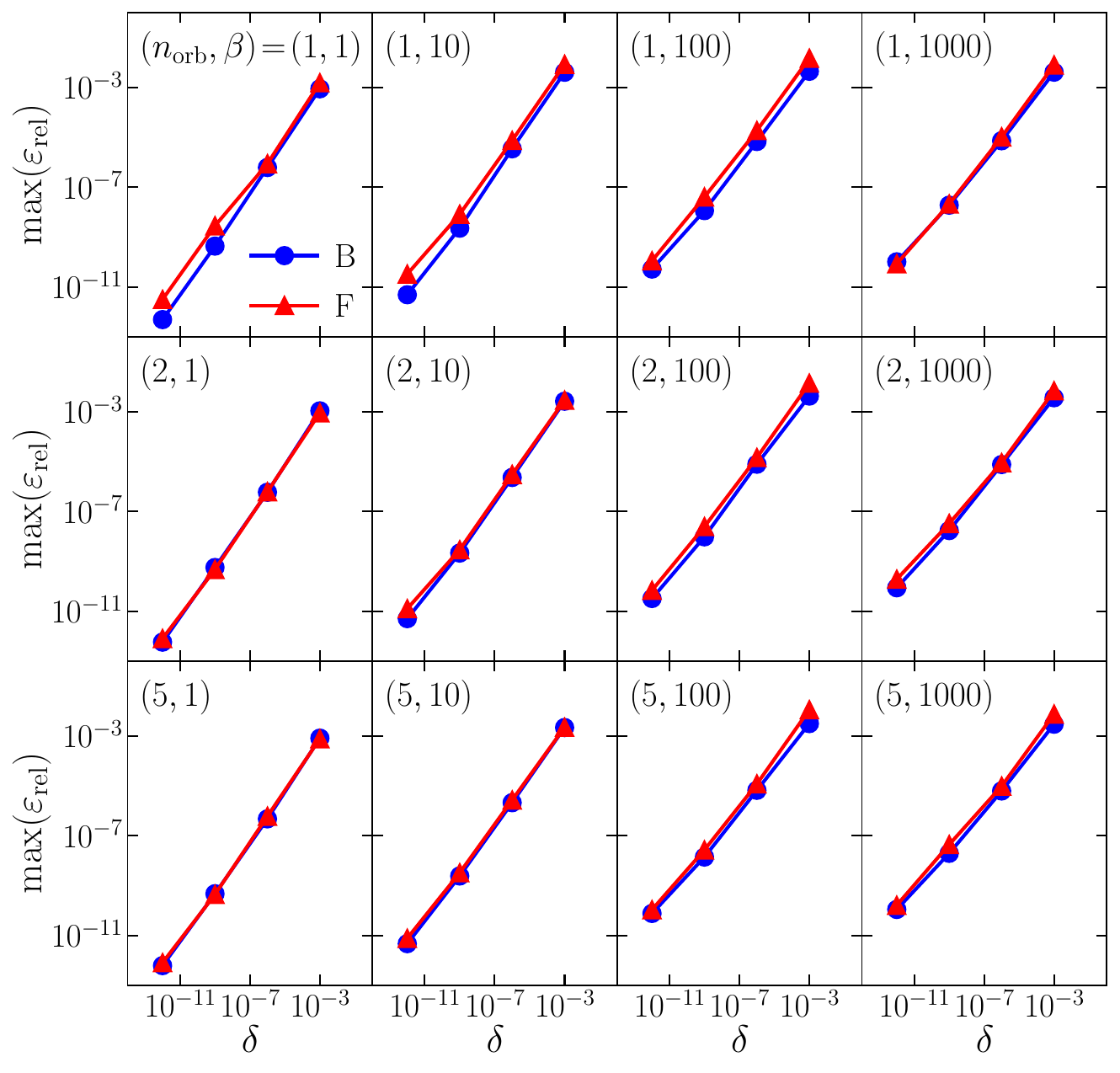}
\caption{The maximum relative deviation \(\varepsilon_{\rm rel}\) (Eq.~\ref{eq:epsapproxecactG}) on the imaginary axis (at and between Matsubara points) among all random configurations, for various temperatures \(\beta\in\{1, 10, 100, 1000\}\) and orbital numbers \(n_\text{orb}\in\{1,2,5\}\), as a function of noise $\delta \in \{10^{-3}, 10^{-6}, 10^{-9}, 10^{-12}\}$. Blue circles: bosonic problems. Red triangles: fermionic problems.}
\label{fig:first_esprit_control}
\end{figure}

We further test the robustness of this approximation in Fig.~\ref{fig:first_esprit_control} for random test functions with bosonic and fermionic properties, by choosing analytically known functions and polluting them with random noise. Our procedure is as follows: 
(i) We randomly choose the number of poles $n_p$ between 1 and $(n_{p})_{\rm max}$; (ii) we then randomly generate $n_p$ real poles $\xi_l$ ($1 \leq l \leq n_p$) within the energy window $[-\omega_{\rm max}, \omega_{\rm max}]$; (iii) we randomly assign a matrix-valued weight $\mat{A}_l$ to each $\xi_l$, where for fermions, $\mat{A}_l$ is positive semidefinite, and for bosons, $\text{sign}(\xi_l) \mat{A}_l$ is positive semidefinite; (iv) we normalize $\mat{A}_l$ such that $\sum_l \mat{A}_l = [d_i, d_j^\dag]_\pm$. This yields a random Green's function with correct analytic properties.

We then evaluate the  Matsubara axis data  as $\mat{G}_{\rm exact}(i\omega_n) = \sum_l \mat{A}_l/(i\omega_n - \xi_l)$. To pollute the data with noise, we add a Gaussian distributed perturbation to each matrix element:
\begin{equation}\label{eq:add_noise}
    [\mat{G}_{\rm input} (i \omega_n)]_{ij} = [\mat{G}_{\rm exact} (i \omega_n)]_{ij} [1 + \delta \times N_\mathbb{C}(0, 1)],
\end{equation}
where $N_\mathbb{C}(0, 1)$ represents the complex-valued normal Gaussian distribution.
Finally, we obtain $\mat{G}_{\rm approx}(i\omega_n)$  from the noisy data $\mat{G}_{\rm input}(i\omega_n)$  and compute the relative maximal deviation on the continuous interpolation interval as
\begin{equation}\label{eq:epsapproxecactG}
    \varepsilon_{\rm rel} = \frac{\max\limits_{\omega_{n_0} \leq y \leq \omega_{n_\omega - 1}}\Vert\mat{G}_{\rm approx}(iy) - \mat{G}_{\rm exact}(iy)\Vert}{\max\limits_{0 \leq n \leq n_\omega - 1}\Vert \mat{G}_{\rm exact} (i\omega_n) \Vert} \;.
\end{equation}
We repeat this procedure to obtain a statistical sample.

The result of this procedure is illustrated in Fig.~\ref{fig:first_esprit_control}. Fermionic systems are shown red,  bosonic ones  in blue. We choose combinations of inverse temperatures $\beta$ between  $1$ and $1000$; $1$, $2$, and $5$ orbitals for each panel and show $\varepsilon_{\rm rel}$ as a function of the noise levels $\delta \in \{10^{-3}, 10^{-6}, 10^{-9}, 10^{-12}\}$.
For each $(\beta, n_{\rm orb}, \delta)$, we randomly generate $10^4$ configurations $\{\mat{A}_l, \xi_l\}$, using $(n_p)_{\rm max} = 100$ and $\omega_{\rm max} = 10$.

We observe that for all tested temperatures, orbital numbers, noise levels, and random configurations, our method provides imaginary frequency approximations that, over the continuous interval $[i \omega_{n_0}, i \omega_{n_\omega - 1}]$, become systematically more accurate as the noise is reduced. Since Eq.~\eqref{eq:h_k_def} of our analytic continuation procedure relies on evaluating precise integrals over the Matsubara axis, this behavior is a prerequisite for systematically improvable continuations.

\begin{figure}[tbh]
    \centering
\includegraphics[width=1.0 \columnwidth]{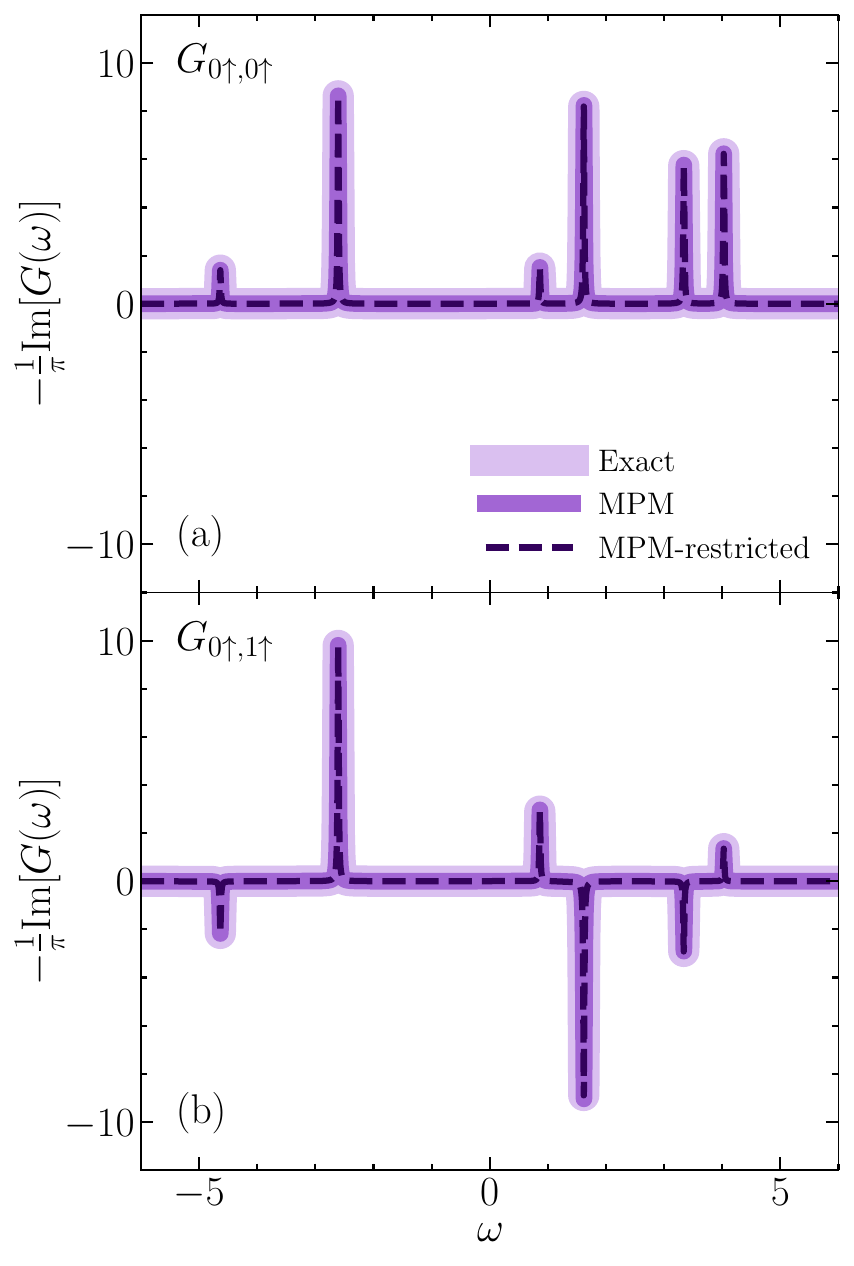}
\caption{Exact solution (light purple, large) and analytic continuation of diagonal (top panel) and off-diagonal (bottom panel) fermionic Matsubara Green's functions of a Hubbard dimer. Continuation is performed using both the mapping (\ref{eq:con_map}) (MPM, purple) and the mapping (\ref{eq:con_map2}) (MPM-restricted, dashed deep purple). The simulations are conducted at \(\beta = 10\).
%\textcolor{red}{this and later plots: consider showing Im G or Re G/Im G as convenient.}[\textcolor{blue}{LZ: done.}]
}
\label{fig:dis_fermi}
\end{figure}
\subsection{Matrix-valued continuations: fermion case}
Next, we show an example of a matrix-valued continuation for a simple multi-orbital system. We analyze the (fermionic) Hubbard dimer given by the Hamiltonian
\begin{align}\label{eq:hubbard_dimer}
    H = H_0 + H_{\rm int} + H_{\rm ext} + H_{\rm sb},
\end{align}
where $H_0 = -\sum_\sigma (t c_{0 \sigma}^\dag c_{1 \sigma} + {\rm h.c.}) - \sum_{i \sigma} \mu n_{i\sigma}$,  $H_{\rm int} = \sum_i U n_{i\uparrow}n_{i\downarrow} - \sum_{i \sigma} \frac{U}{2} n_{i\sigma}$, with $n_{i\sigma} = c_{i\sigma}^\dag c_{i\sigma}$ being the occupation number operator and $i\in \left\{0,1\right\}$. Additional terms $H_{\rm ext} = \sum_i h (n_{i \uparrow} - n_{i\downarrow})$ and $H_{\rm sb} = U_a (n_{0\uparrow}n_{0\downarrow} - n_{1\uparrow}n_{1\downarrow}) + \mu_a (n_{0\uparrow} + n_{0\downarrow} - n_{1\uparrow} - n_{1\downarrow}) + h_a (n_{0\uparrow} - n_{0\downarrow} - n_{1\uparrow} + n_{1\downarrow})$ are added to break symmetries and generate additional excitations. This system has two orbitals (four spin-orbitals), and an analytic solution for the spectral function is available. The Hubbard dimer is a starting point for studying the extended Hubbard model which is a paradigmatic model of strongly correlated electron systems \cite{Arovas22,Qin22}.

Since the energy levels are discrete, sharp resonances, extracting accurate spectral functions using the maximum entropy method is known to be difficult \cite{Fei21_Cara}.

To illustrate the matrix-valued continuation, we study the system in the absence of noise and evaluate all Green's function data 
on the Matsubara axis in double precision.
The parameters are chosen to match those in Ref.~\cite{Fei21_Cara}, with $t=1$, $U=5$, $\mu=0.7$, $h=0.3$, $U_a=0.5$, $\mu_a=0.2$, and $h_a=0.03$. We perform exact diagonalization at $\beta = 10$ to obtain both the exact spectral function and the exact Matsubara Green's function $\mat{G}(i \omega_n)$.

The Matsubara data is then processed using the methodology of Sec.~\ref{sec:method}, with both the generic holomorphic mapping (\ref{eq:con_map}), which allows poles anywhere on the complex plane, and the restricted mapping (\ref{eq:con_map2}), which restricts them to the real axis.

Fig.~\ref{fig:dis_fermi} shows that both methods successfully recover the correct diagonal (top panel) and off-diagonal (bottom panel) results for the Hubbard dimer. The broadening parameter, $\eta$, is set to 0.01 for visualization purposes. The recovered spectral functions share the same set of poles, with matrix-valued weights being positive semidefinite.

\begin{figure}[bth]
    \centering
\includegraphics[width=1.0 \columnwidth]{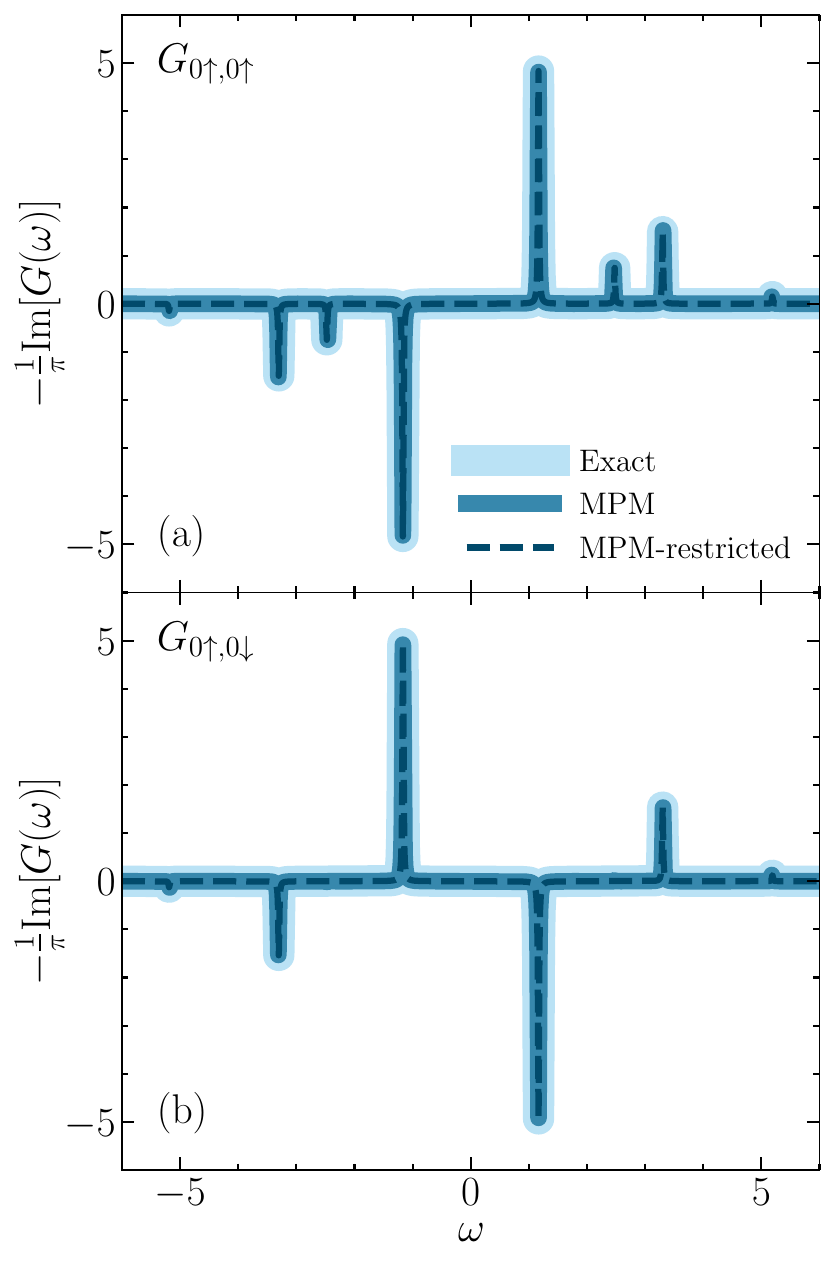}
\caption{Analytic continuation of bosonic Matsubara Green's functions obtained from the exact diagonalization of the same discrete Hubbard dimer system as in Fig.~\ref{fig:dis_fermi}.}
\label{fig:dis_boson}
\end{figure}
\subsection{Matrix-valued continuations: bosonic case}

In Fig.~\ref{fig:dis_boson}, we obtain similar results for a discrete bosonic response function, using the Hamiltonian of Eq.~(\ref{eq:hubbard_dimer}), but replacing the fermionic operators $d$ and $d^\dag$ in Eq.~(\ref{eq:green_func}) with bosonic density operators $n_{i \sigma} = n^\dag_{i \sigma} = c_{i\sigma}^\dag c_{i\sigma}$. All other parameters remain the same, except for $U$, which is now set to 2.5 for better visualization. As shown in Fig.~\ref{fig:dis_boson}, both restricted and unrestricted methods successfully recover the correct spectra.

Note that while it is straightforward to obtain bosonic continuations with the method of Sec.~\ref{sec:method}, maximum entropy-related methods struggle for bosonic functions due to the conditioning of the bosonic kernel  and the fast decay of bosonic response functions \cite{Jarrell96}. For noise-free data, a generalization of the bosonic scalar Nevanlinna method \cite{Nogaki23} to matrix-valued Carath\'{e}odory functions may also be used.

\begin{figure}[tbh]
    \centering
\includegraphics[width=1.0 \columnwidth]{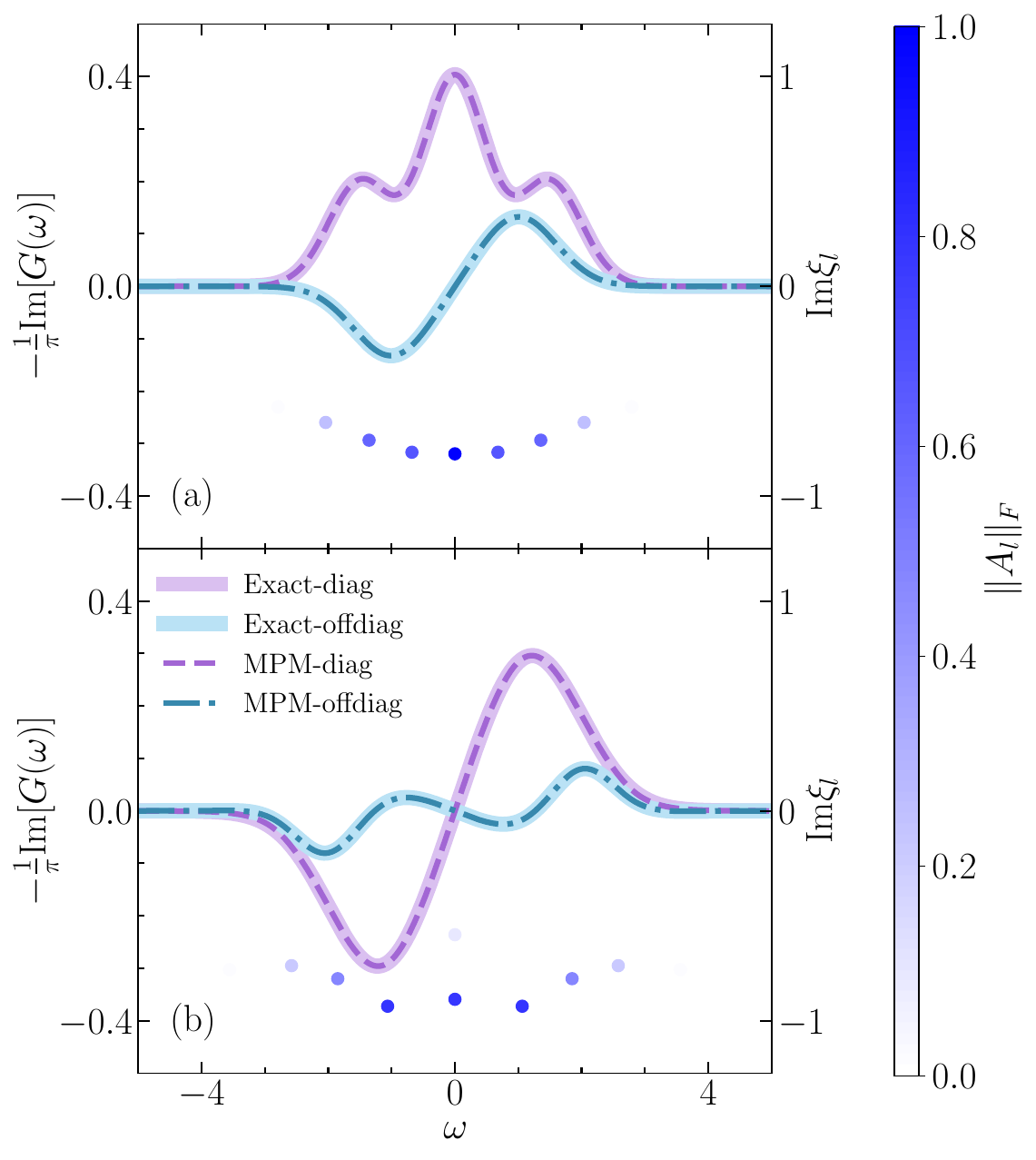}
\caption{Analytic continuation of a continuous spectral function. Top panel: fermionic case; bottom panel: bosonic case. Blue circles represent the locations of the shared poles, with their weights indicated by the color bar. The simulations are carried out at \(\beta = 20\).}
\label{fig:cont_fermi_boson}
\end{figure}

\begin{figure}[tbh]
    \centering
\includegraphics[width=1.0 \columnwidth]{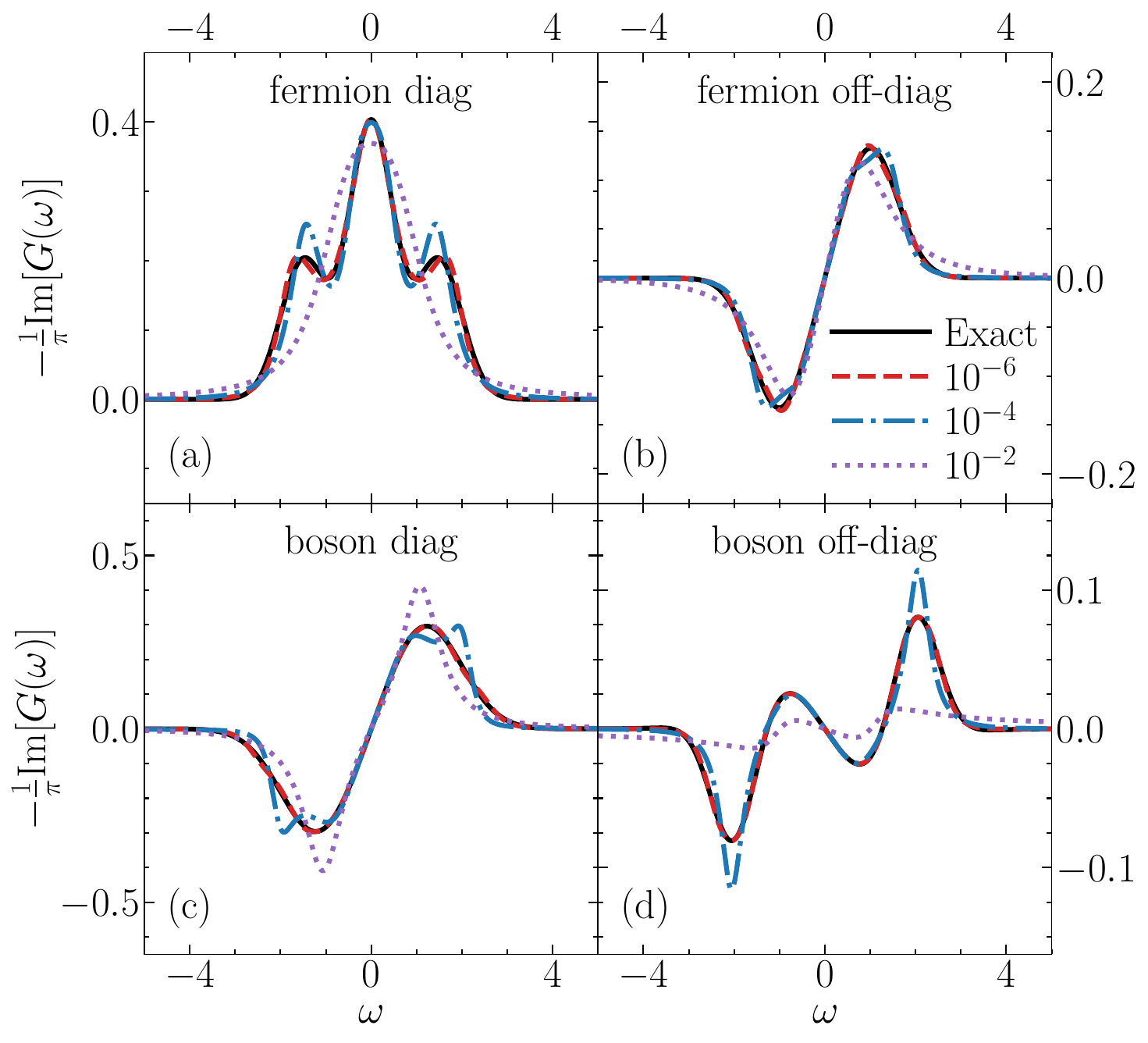}
\caption{Noise robustness analysis of the MPM algorithm, performed by adding Gaussian random noise with amplitudes $\delta \in \{10^{-2}, 10^{-4}, 10^{-6}\}$. (a) Diagonal and (b) off-diagonal fermionic case. (c) Diagonal and (d) off-diagonal bosonic case.}
\label{fig:cont_noise_robust}
\end{figure}

\subsection{Matrix-valued continuations: continuous system}

In Fig.~\ref{fig:cont_fermi_boson}, we examine the method for the case of a continuous spectral function. Poles of this approximation have to be in the lower half of the complex plane, since poles on the real axis would result in a discrete spectral function. 

For the fermion case, we model the diagonal and off-diagonal spectra as  $A_{\rm diag}(\omega) = 0.25 g(\omega, -1.5, 0.5) + 0.5 g(\omega, 0, 0.5) + 0.25 g(\omega, 1.5, 0.5)$ and $A_{\rm off-diag}(\omega) = -0.2 g(\omega, -1, 0.6) + 0.2 g(\omega, 1, 0.6)$, where $g(\omega, \mu, \sigma) = \frac{1}{\sqrt{2\pi} \sigma} \exp\left(-\frac{(\omega - \mu)^2}{2\sigma^2}\right)$ is the Gaussian function. For the bosonic case, we use $A_{\rm diag}(\omega) = -0.6 g(\omega, -1.2, 0.8) + 0.6 g(\omega, 1.2, 0.8)$ and $A_{\rm off-diag}(\omega) = -0.13 g(\omega, -2, 0.5) + 0.1 g(\omega, -1, 1) - 0.1 g(\omega, 1, 1) + 0.13 g(\omega, 2, 0.5)$. The Matsubara Green's functions are calculated from Eq.~(\ref{eq:Mat_spec}) at $\beta = 20$. The matrix-valued Matsubara data are then processed  with the mapping (\ref{eq:con_map}).

These superpositions of Gaussians are not easily approximated by sums of poles, as the pole spectra correspond to Lorentzians. Nevertheless, both fermionic and bosonic spectra are recovered well with shared poles.
Fig.~\ref{fig:cont_fermi_boson} shows the diagonal  (thick purple lines) and off-diagonal (thick blue lines) spectra along with the recovered analytic continuation of the matrix-valued problem (dashed purple and dot-dashed blue lines). The fermion problem is shown in the top graph, the bosonic problem in the bottom graph, both evaluated at $\eta=0^+$. Also shown are the pole locations $\xi_l$ and the Frobenius norm of the pole strength, as indicated in the color bar on the right. Note that poles are shared between diagonal and off-diagonal Green's function components.

Fig.~\ref{fig:cont_noise_robust} analyzes the sensitivity of the analytic continuation to random noise. We evaluate the robustness to noise by adding Gaussian random noise $\delta \in \{10^{-2}, 10^{-4}, 10^{-6}\}$ to the input Matsubara, as described in Eq.~(\ref{eq:add_noise}). In agreement with the scalar-valued case \cite{zhang2024minimal}, the main features of the spectra are well reproduced even with relatively high noise levels. As the noise level decreases, both the diagonal and off-diagonal spectra rapidly converge to the exact result.

\subsection{Application to scalar and matrix-valued real-materials data}\label{sec:realistic}
\begin{figure*}[bth]
    \centering
    \includegraphics[width=2 \columnwidth]{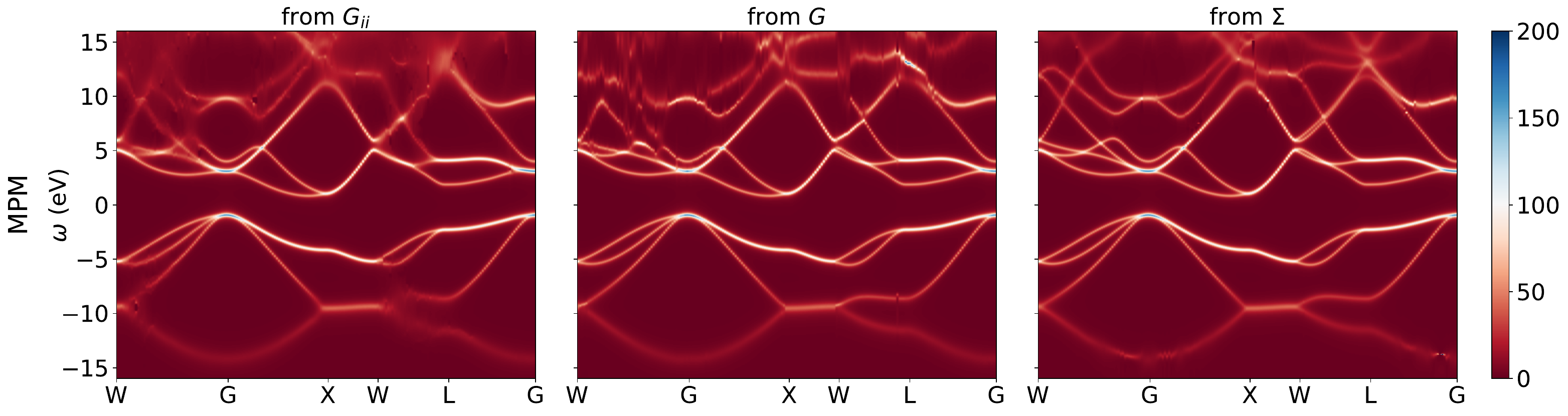}
    \includegraphics[width=2 \columnwidth]{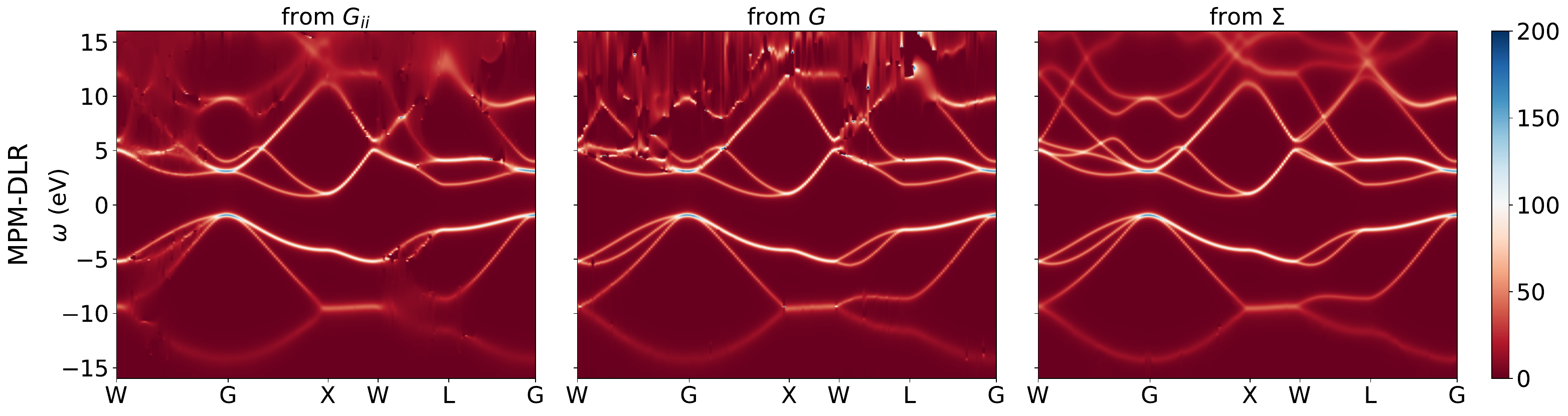}
    \includegraphics[width=2 \columnwidth]{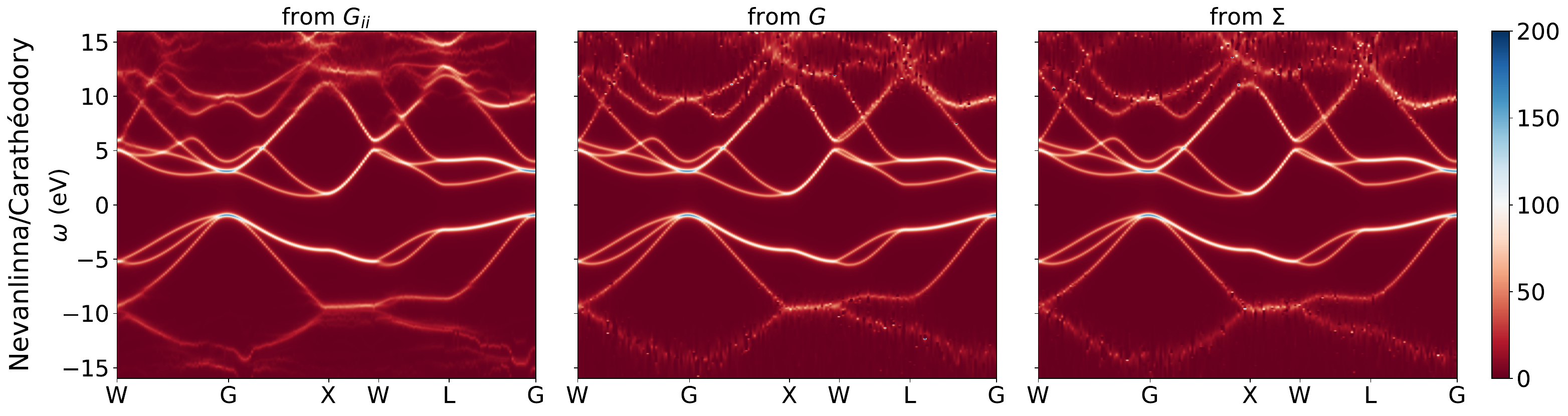}
\caption{Spectral function (band structure) of Si obtained by analytic continuation of self-consistent GW data \cite{Yeh22,Fei21_Cara} at inverse temperature $\beta = 700 \;{\mathrm{Ha}}^{-1}$. We use the same input Matsubara data for all plots and, for clearer visualization, evaluate the spectral function at $\eta = 0.005$ above the real axis, introducing an artificial broadening. Left panel: scalar continuation of the diagonal elements of the Green's function. Middle panel: matrix-valued continuation of the Green's function. Right panel: matrix-valued continuation of the self-energy followed by a Dyson equation on the real axis to obtain the real frequency Green's function. Top set of plots: continuation with the method presented in Sec.~\ref{sec:method}. Middle set of plot: evaluation with DLR coefficients as presented in Sec.~\ref{sec:dlr}. Bottom panel: evaluation with Nevanlinna \cite{Fei21} and Carath\'eodory \cite{Fei21_Cara} methods.}
\label{fig:band}
\end{figure*}
As a final example, we present an application of the method to a real materials simulation. Response functions in such simulations typically spread a large range of energies and are therefore very sensitive to inaccuracies in the real- or imaginary-frequency grids and the continuation method.

We choose crystalline Si as an example system, simulated in the fully self-consistent GW approximation with the GREEN open-source software package~\cite{ISKAKOV2025109380}. Simulations are carried out in Gaussian orbitals (basis set gth-dzvp-molopt-sr~\cite{vandevondele2007gaussian} with gth-pbe pseudopotential~\cite{goedecker1996separable}), with integrals generated by the PySCF package~\cite{sun2020recent}. Calculations are performed on a $6 \times 6 \times 6$ $k$-point grid with 26 orbitals per unit cell. After achieving convergence, the orbitals are interpolated on a grid with 200 $k$ points along a high-symmetry path in reciprocal space and then transformed into an orthogonal orbital basis. The simulations are conducted at $\beta = 700 \;{{\mathrm{Ha}}^{-1}}$, where $\mathrm{Ha}$ denotes the Hartree energy. To perform the continuation, the Matsubara data are evaluated~\cite{Li20} on the first 1000 positive Matsubara frequencies from 104 IR frequencies~\cite{shinaoka2017compressing}.

We present results of the continuation presented in this paper in the top panel of Fig.~\ref{fig:band}. The top left panel displays the sum of the scalar analytic continuation of the diagonal elements. In this approach, each continuation is carried out as described in Ref.~\cite{zhang2024minimal}, with $n_0$ automatically determined using the method outlined in Sec.~\ref{sec:firstProny}. The middle panel presents results from a matrix-valued continuation where the Green's functions share common poles, and all diagonal and off-diagonal entries are simultaneously continued, respecting the positive semidefinite pole structure. Finally, the right panel shows results from the matrix-valued continuation of the self-energy, followed by the evaluation of the Dyson equation on the real axis to obtain the real-frequency Green's function. This procedure is analogous to the one described in Ref.~\cite{Fei21_Cara}. The poles are allowed to lie anywhere in the lower half of the complex plane.

Results from the accelerated algorithm using DLR coefficients for diagonal, matrix, and self-energy continuation are shown in the middle row of Fig.~\ref{fig:band}.

In the lower panel of Fig.~\ref{fig:band}, we present results from state-of-the-art methods: the Nevanlinna continuation~\cite{Fei21} for scalar-valued cases and the Carathéodory continuation~\cite{Fei21_Cara} for matrix-valued cases, both performed directly on the IR grid. The broadening parameter $\eta$ is set to 0.005 for all plots to ensure a consistent comparison.

All continuation methods result in sharp bands and consistent band gaps; the degeneracies of the bands at high-symmetry points and along high-symmetry directions are accurately recovered. Green's function and self-energy continuations agree on the value of band gaps and the main spectral features. High-lying bands are generally sharper in the continuations of the self-energy. The DLR based algorithm presented in Sec.~\ref{sec:dlr} typically generates a less clear band structure than the algorithm employing the method of Sec.~\ref{sec:firstProny} for both the scalar and the matrix continuation of the Green's function. In contrast, when continuing from the self-energy, the two methods exhibit nearly identical resolution, which is observed to be superior to that of the Green's function continuation and comparable to results from the Nevanlinna or Carathéodory continuation.

All  results from the minimal pole method and its DLR variant are obtained using the mapping in Eq.~(\ref{eq:con_map}). Due to the sharp features present in the band structure, one can alternatively use the mapping in Eq.~(\ref{eq:con_map2}) to restrict the poles to the real axis when continuing scalar and matrix-valued Green's functions. %This approach tends to provide better resolution. %\del{However, visualization then introduces additional broadening effects.} \add{However, the spectra must be evaluated at a finite \(\eta\) instead of \(0^+\), and as a result, the peak widths can no longer be reliably estimated.}

Compared to the Carath\'{e}odory continuation method discussed in Ref.~\cite{Fei21_Cara}, which interpolates a causal matrix-valued solution, the fitting approach presented here operates in double precision and is robust to noise. Due to the need for multi-precision arithmetic, results for the Carathéodory continuation as implemented 
by the TRIQS Nevanlinna package~\cite{ISKAKOV2024109299} [bottom panel of Fig.~\ref{fig:band}] required ten times more CPU time than the results based on the method introduced in Sec.~\ref{sec:firstProny} [top panel of Fig.~\ref{fig:band}]. Furthermore, the continuation based on the DLR expansion, presented in Sec.~\ref{sec:dlr} [middle panel of Fig.~\ref{fig:band}], completed in minutes on a single core, resulting in an additional computational speedup of three orders of magnitude.

\section{Conclusions}\label{sec:conclusions}
The analytic continuation of matrix-valued correlation functions of data from finite-temperature field theories is essential for interpreting spectral functions in systems with off-diagonal correlations, for evaluating derived quantities such as optical response functions on the real axis, and for interpreting anomalous order in superconducting systems.

This paper presents a theoretical framework and numerical method based on the construction of an approximation with shared poles for analyzing and continuing such correlation functions. We show that the method works well in a variety of setups, including fermionic and bosonic, discrete and continuous, as well as model and real materials systems. Unlike methods based on interpolation with causal functions \cite{Fei21_Cara}, this method is robust to noise and eliminates the need for multi-precision arithmetic. In contrast to methods based on smoothness principles \cite{Jarrell96}, it is capable of resolving both sharp and smooth features.

Numerical analytic continuation remains an ill-posed problem. In practice, whether the minimum information principle imposed by the pole approximation presented here leads to a suitable spectral function for a system at hand needs to be evaluated on a case-by-case basis.

The representation and compression of spectral functions on the real axis with a minimum number of complex poles using the methods presented here, and the solution of dissipative systems based on these pole representations are closely related problems that will be examined in future work.

Our paper is accompanied by an open-source software implementation of all methods presented here, written in the programming language Python~\cite{supp,zhang_2024_13936894,MiniPoleCode}.
\begin{acknowledgments}
We thank Chia-Nan Yeh, Sergei Iskakov, Andre Erpenbeck and Selina Dirnb\"{o}ck for helpful discussions. This material is based upon work supported by the National Science Foundation under Grant No.~2310182. Results made use of the Green \cite{Green,ISKAKOV2025109380}, irbasis \cite{CHIKANO2019181}, libdlr \cite{kaye2022libdlr} and TRIQS Nevanlinna \cite{ISKAKOV2024109299} open source software packages.
\end{acknowledgments}

\bibliography{reference}

\end{document}